\documentclass[runningheads]{llncs}
\usepackage{graphicx}
\usepackage{subfigure}
\usepackage{booktabs} 
\usepackage{amsmath}
\usepackage{bbm}
\usepackage{array}
\usepackage{wrapfig}

\usepackage{fontawesome}
\usepackage{multirow}
\usepackage{float}

\begin{document}

\title{Towards Confident Detection of Prostate Cancer using High Resolution Micro-ultrasound}
%
%
\author{Mahdi Gilany\textsuperscript{1(\faEnvelopeO)}, Paul Wilson\textsuperscript{1}, Amoon Jamzad\textsuperscript{1}, Fahimeh Fooladgar\textsuperscript{2}, Minh Nguyen Nhat To\textsuperscript{2}, Brian Wodlinger\textsuperscript{3}, Purang Abolmaesumi\textsuperscript{2}, Parvin Mousavi\textsuperscript{1}}

%
\authorrunning{M. Gilany et al.}
\titlerunning{Towards Confident Prostate Cancer Detection}
%
\institute{\textsuperscript{1}School of Computing, Queen's University, Kingston, Canada \\
\textbf{mahdi.gilany@queensu.ca}\\
\textsuperscript{2}Department of Electrical and Computer Engineering, University of British
Columbia, Vancouver, BC, Canada\\
\textsuperscript{3}Exact Imaging, Markham, Canada}
\maketitle              
\begin{abstract}
MOTIVATION:
Detection of prostate cancer during transrectal ultrasound-guided biopsy is challenging. The highly heterogeneous appearance of cancer, presence of ultrasound artefacts, and noise all contribute to these difficulties. Recent advancements in high-frequency ultrasound imaging - micro-ultrasound - have drastically increased the capability of tissue imaging at high resolution. Our aim is to investigate the development of a robust deep learning model specifically for micro-ultrasound-guided prostate cancer biopsy. For the model to be clinically adopted, a key challenge is to design a solution that can confidently identify the cancer, while learning from coarse histopathology measurements of biopsy samples that introduce weak labels. METHODS: We use a dataset of micro-ultrasound images acquired from 194 patients, who underwent prostate biopsy. We train a deep model using a co-teaching paradigm to handle noise in labels, together with an evidential deep learning method for uncertainty estimation. We evaluate the performance of our model using the clinically relevant metric of accuracy vs. confidence. RESULTS: Our model achieves a well-calibrated estimation of predictive uncertainty with area under the curve of 88$\%$. The use of co-teaching and evidential deep learning in combination yields significantly better uncertainty estimation than either alone. We also provide a detailed comparison against state-of-the-art in uncertainty estimation.


\end{abstract}
\section{Introduction}

Prostate cancer (PCa) is the second most common cancer in men worldwide~\cite{smith2018canadian}. The standard of care for diagnosing PCa is histopathological analysis of tissue samples obtained via systematic prostate biopsy under trans-rectal ultrasound (TRUS) guidance. TRUS is used for anatomical navigation rather than cancer targeting. The appearance of cancer on ultrasound is highly heterogeneous and is further affected by imaging artifacts and noise, resulting in low sensitivity and specificity in PCa detection based on ultrasound alone. 





Substantial previous literature and large multi-center trials report low sensitivity of systematic TRUS biopsy. In \cite{ahmed2017diagnostic}, authors compare diagnostic accuracy of TRUS biopsy and multi-parametric MRI (mp-MRI). They report sensitivity of systematic TRUS biopsy as low as 42-55\% compared to 88-96\% for mp-MRI. However, they report low specificity of 36-46\% for mp-MRI compared to 94-98\% for TRUS. 

Fusion of mp-MRI imaging with ultrasound can enable targeted biopsy by identifying cancerous lesions in the prostate~\cite{rai2021magnetic,siddiqui2013magnetic}. Fusion biopsy involves either manual or semi-automated registration of lesions identified in mp-MRI with real-time TRUS. This process can be time-consuming and inaccurate due to registration errors and patient motion. It is therefore highly desirable to improve the capability of biopsy targeting using ultrasound imaging alone at the point of care.

The recent development of high frequency ``micro-ultrasound" technology  allows for the visualization of tissue at higher resolution than conventional ultrasound. A qualitative scoring system based on visual analysis of micro-ultrasound images called  the PRI-MUS (prostate risk identification using micro-ultrasound) protocol~\cite{ghai2016assessing} has been proposed to estimate PCa likelihood. Several studies have shown that micro-ultrasound can detect PCa with sensitivity comparable to that of mp-MRI using this grading system~\cite{abouassaly2020impact,eure2019comparison}. A recent systematic review and meta analysis analyzing 13 published studies with 1,125 total participents found that micro-ultrasound guided prostate biopsy and mp-MRI imaging targeted prostate biopsy resulted in comparable detection rates for PCa \cite{sountoulides2021micro}.  
Research on this technology is in early stages and relatively few quantitative methods are reported. Rohrbach et al.~\cite{rohrbach2018high} use a combination of manual feature selection with machine learning as the first quantitative approach to this problem. Shao et al.~\cite{shao2020improving} use a deep learning strategy with a three-player minimax game to tackle data source heterogeneity. While these studies show significant potential of micro-ultrasound as a diagnostic tool for PCa, methods to-date primarily focus on improving accuracy for cancer prediction. We argue that in addition, confidence in detection of cancer can play a significant role for adoption of this technology to ensure that predictions can be clinically trusted. Towards this end, we propose to address several key challenges.


Machine learning models built from ultrasound data rely on ground truth labels from histopathology that are coarse and only approximately describe the spatial distribution of cancer in a biopsy core~\cite{javadi2021training,rohrbach2018high,shao2020improving}. The lack of finer labels cause two challenges: first, labels assigned to patches of ultrasound images in a biopsy core may not match the ground truth tissue, resulting in weak labels; second, biopsies include other types of tissue such as fibromuscular cells, benign prostatic hyperplasia and precancerous changes. Many of these tissues are unlabeled in a histopathology report, which will result in out-of-distribution (OOD) data. Therefore, effective learning models for micro-ultrasound data should be robust to label noise and OOD samples.

Several solutions have been presented to address the above issues, mainly by quantifying the uncertainty of predictions~\cite{ABDAR2021243,lakshminarayanan2017simple,gal2016dropout}. Predictive uncertainty can be used as a tool to discard unreliable and OOD samples. Evidential deep learning (EDL)~\cite{sensoy2018evidential} and ensemble methods~\cite{lakshminarayanan2017simple} are amongst such approaches. In particular, evidential learning is computationally light, run-time efficient and theoretically grounded, hence it fits our clinical purpose here. Learning from noise in labels (i.e. weak labels) has also been addressed before using methods that 1) estimate noise; 2) modify the learning objective function, or 3) use alternative optimization~\cite{han2020survey}. Among these, co-teaching~\cite{han2018co} has been shown to be a successful baseline that can be easily integrated with any uncertainty quantification method. 

In this paper, for the first time, we propose a learning model for PCa detection using micro-ultrasound that can provide an estimate of its predictive confidence and is robust to weak labels and OOD data. We address label noise using co-teaching and utilize evidential learning to estimate uncertainty for OOD rejection, resulting in confident detection of PCa. We assess our approach by examining the classification accuracy and uncertainty calibration (i.e. the tendency of the model to have high levels of certainty on correct predictions). We compare our methodology to a variety of uncertainty methods with and without co-teaching and demonstrate significant improvements over baseline. We show that applying an adjustable threshold to discard uncertain predictions yields great improvements in accuracy. By allowing correct and confident predictions, our approach could provide clinicians with a powerful tool for computer-assisted cancer detection from ultrasound.

\section{Materials and Methods}
\subsection{Data}
\label{section:data}
Data is obtained from 2,335 biopsy cores of 198 patients who underwent transrectal ultrasound-guided prostate biopsy through a clinical trial and after institutional ethics approval is provided. A 29~MHz micro-ultrasound system and transducer (ExactVu, Markham) was used for data acquisition. A single sagittal ultrasound image composed of 512 lateral radio frequency (RF) lines was obtained prior to the firing of the biopsy gun for each core. Primary and secondary Gleason grades, together with an estimate of the fraction of cancer relative to the total core area (the so-called ``involvement of cancer") are also provided for each patient. We under-sampled benign cores in order to obtain an equal proportion of cancerous and benign cores during training and evaluation, resulting in 300 benign and 300 cancerous cores, respectively. As in \cite{rohrbach2018high}, we exclude cores with involvement less than 40\% to learn from data that better represents PCa. We hold out the data from 27 patients as a test set, with the remaining 161 used for training and cross-validation. 

\subsubsection{Pre-processing:}

\begin{figure}[t]
    \centering
    \includegraphics[scale=0.5]{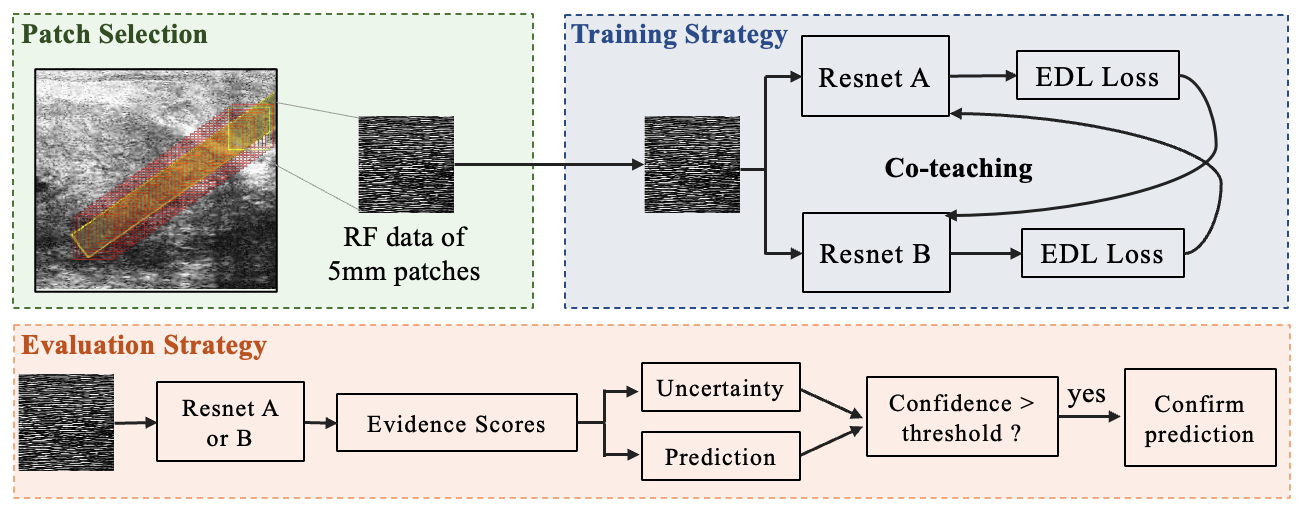}
    \caption{Top left: Patches are extracted from the needle region. Top Right: During training, ``clean" examples are selected by the peer model for training updates. Bottom: The model predicts evidence scores which are used to calculate predictions and uncertainty. Predictions with high uncertainty are rejected. }
    \label{fig:my_label}
\end{figure}

For each RF ultrasound image, a rectangular region of interest (ROI) corresponding to the approximate needle trace area is determined by using the angle and location of the probe-mounted needle relative to the imaging plane (Fig.~\ref{fig:my_label}, yellow region). This ROI is intersected with a manually drawn prostate segmentation mask to exclude non-prostatic tissue. Overlapping patches are extracted corresponding to 5~\textit{mm} $\times$ 5~\textit{mm} tissue regions with an overlap of 90\% covering the ROI. These patches are up-sampled in the lateral direction and down-sampled in the axial direction by factors of 5 to obtain a uniform physical spacing of pixels in both directions. This results in a patch of 256 by 256 pixels. Ultrasound data in each patch are normalized to a mean of 0 and standard deviation of 1. Patches are assigned a binary label of 0 (benign) or 1 (cancerous) depending on the pathology of the core. The patches and their associated labels are inputs to our learning algorithms. 


\subsection{Methodology}

We propose a micro-ultrasound PCa detection learning model that is robust to challenges associated with weak labels and OOD samples. In this section, we first define the problem followed by descriptions of co-teaching as a strategy for dealing with weak labels. Next, we incorporate evidential deep for quantifying prediction uncertainty and excluding suspected OOD data. Finally, we present evaluation metrics to assess our methods.

\subsubsection{Weak Labels and OOD:}
Let $X_i = \{x_1, x_2, ..., x_{n_i}\}$ refer to a biopsy core where $n_i$ number of patches extracted from needle region (Figure \ref{fig:my_label}). For each biopsy core $X_i$, pathology reports a label $Y_i$ and the length of cancer $L_i$ in core, which is a rough estimate between zero and the biopsy sample length. Following previous work in PCa detection~\cite{rohrbach2018high,javadi2021training}, we assign coarse pathology labels $Y_i$ to all extracted patches $\{x_1, x_2, ..., x_{n_i}\}$ due to the lack of finer patch-level labels. Therefore, many assigned labels to patches may not necessarily match with the ground truth and they are inherently weak. Additionally, other tissue than cancer, present in the core, does not have any gold standard labels. Therefore, there is also OOD data. 
\subsubsection{Co-teaching:} 
We propose to use a state-of-the-art method, co-teaching, to address label noise for micro-ultrasound data~\cite{han2018co}. For weak label methods, we rely on the findings of \cite{javadi2021training} showing the success of co-teaching method, and \cite{toincreasing}, which found that co-teaching significantly out-performed other methods such as robust loss functions.
This approach simultaneously trains two similar neural networks with different weight initializations. According to the theory of co-teaching, neural networks initially learn simpler and cleaner samples then overfit to noisy input. Therefore, during each iteration, each network picks a subset of samples with lower loss values as potentially clean data and trains the other network with those samples. In a batch of data with size $N$, only $R(e)*N$ number of samples are selected by each network as clean samples, where $R(e)$ is the ratio of selection starting from $1$ and gradually decreasing to a fixed value $1-\gamma$. Formally we have $R(e)=1-\min(\frac{e}{e_{\max}}, \gamma)$, where $\gamma \in [0, 1]$ is a hyper-parameter, and $e$ and $e_{\max}$ are the current and maximum number of epochs, respectively. Using two networks prevents confirmation bias from arising.

\subsubsection{Evidential Deep Learning:}

Evidential deep learning (EDL) \cite{sensoy2018evidential} uses the concepts of \emph{belief} and \emph{evidence} to formalize the notion of uncertainty in deep learning.
A neural network is used to learn the parameters of a prior distribution for the class likelihoods instead of point estimates of these likelihoods. Given a binary classification problem where $P(y = 1 | x_i) = p_i$, instead of estimating $p_i$, the network estimates parameters $e_0, e_1$ such that $p_i \sim \text{Beta}(e_0 + 1, e_1 + 1)$. These parameters are then referred to as evidence scores for the classes, and used to generate a belief mass and uncertainty assignment, via $b_0 = \frac{e_0}{S}, b_1 = \frac{e_1}{S}, U = \frac{2}{S} $, where $S = \sum_{i=0}^{1} e_i + 1$. Note that $b_0 + b_1 + U = 1$. $U$ ranges between $0$ and $1$ and is inversely proportional to our overall level of belief or evidence for each class. It is worth mentioning that term confidence is also used often instead of uncertainty with confidence being $1-U$. 

The network is trained to minimize an objective function based on its Bayes Risk as an estimator of the likelihoods ${p_i}$. If the network produces evidences $e_0, e_1$ for sample $i$, the loss and predicted uncertainty for this sample are
\begin{equation}
\label{unc_eq}
   \mathcal{L}_i = \sum_{i=1}^{n} E_{p_i \sim \text{Beta}(e_0 + e_1)}\big(|p_i - y_i|^2\big), U_i = \frac{2}{e_0+e_1+2} ,
\end{equation}
where $e_0$ and $e_1$ are the network outputs. The loss also incorporates a KL divergence term, which encourages higher uncertainty on predictions that do not contribute to data fit. The method offers a combination of speed (requiring only a single forward pass for inference) and well-calibrated uncertainty estimation with a solid theoretical foundation.

\subsubsection{Clinical Evaluation Metrics:} 


The goal of our model is to provide the operator with clinically relevant information, such as real-time identification of potential biopsy targets. It should also state the degree of confidence in its predictions such that the operator can decide when to accept the model's suggestions or defer to their own experience. To measure these success criteria, we propose several evaluation metrics. 

Accuracy reported at the level of patches (the basic input to the model) can be misleading due to weak labeling (some correct predictions are recorded as incorrect because of incorrect labels). Therefore, we propose accuracy reported at the level of biopsy cores as a more relevant alternative. We determine core-based accuracy using core-wise predictions aggregated from patch-wise predictions for the core. Specifically, the average of patch predicted labels is used as a probability score that cancer exists in the core \cite{toincreasing,to2022coarse}. To model uncertainty at the core level, patch-wise predictions that do not meet a specified confidence threshold are ignored when calculating this score, and if more than 40$\%$ of the patch predictions for a core fall below this threshold, the entire core prediction is considered ``uncertain". 

We also use ``uncertainty calibration", a metric that assesses how accurate and representative the predicted uncertainty or confidence is (in terms of true likelihood). To compute calibration, we compute Expected Calibration Error (ECE)~\cite{guo2017calibration}, which measures the correspondence between predictive confidence and empirical accuracy. ECE is calculated by grouping the predictions so that each prediction falls into one of the $S$ equal bins produced between zero and one based on its confidence score: 
\begin{equation}
\label{ece}
\begin{split}
    \text{ECE} & = \sum_{s=1}^{S}{\frac{n_s}{N}|\text{acc}(s)-\text{conf}(s)|} ,
\end{split}
\end{equation}
where $S$ denotes the number of bins (10 used in this paper), $n_s$ the number of predictions in bin $s$, $N$ the total number of predictions, and acc$(s)$ and conf$(s)$ the relative accuracy and average confidence of bin $s$, respectively.

\section{Experiments and Results}
\begin{table}[t]
  \caption{Effect of co-teaching on accuracy and calibration error.}
  \label{coteaching_table}
  \centering
  \begin{tabular}{l c c c c c c}
    \toprule
    \textbf{Method}     & AUC   & Sensitivity & Specificity & Patch\,B-accuracy & 
    ECE \\
    \midrule
    \multirow{2}{10em}{EDL} &  \textbf{88.27} & 71.32 & 84.80 & 67.47 & 
    0.1989\\
    & \scriptsize{$\pm$ 2.66} & \scriptsize{$\pm$ 1.23} & \scriptsize{$\pm$ 7.01} & \scriptsize{$\pm$ 2.47} & \scriptsize{$\pm$ 0.0142}\\
    
    \\
    
    \multirow{2}{10em}{EDL + Co-teaching (ours)} & \textbf{87.76} & 67.38 & 88.20 & \textbf{71.25} & 
    \textbf{0.1379} \\
    & \scriptsize{$\pm$ 1.82} & \scriptsize{$\pm$ 4.91} & \scriptsize{$\pm$ 6.85} & \scriptsize{$\pm$ 1.16} & \scriptsize{$\pm$ 0.0258}\\
    
    \bottomrule
  \end{tabular}
\end{table}
From all data, 161 patients (392 cores, 12664 patches) are used for training and a further 40 patients are used as a validation set for model selection and tuning. We hold out a set of randomly selected, mutually exclusive, patients as test set (27 patients, 80 cores, 2808 patches).  All experiments, except for the ensemble method, are repeated nine times with three different validation sets, each with three different initializations; the average of all runs is reported. For the ensemble method, as suggested in~\cite{lakshminarayanan2017simple}, five different models with different initialization are used for estimating true prediction probabilities, $p(y_i|x_i)$. This process is done with five different validation sets, resulting in a total of 25 runs. As a backbone network, we modify ResNet18~\cite{he2016deep} by using only half of the layers in each residual block. We found this reduction in layers to improve model performance, likely by reducing overfitting. Two copies of modified ResNet with different initializations are used for the co-teaching framework. For our choice of $\gamma$, we emprically found 0.4 to be the best. We employ the NovoGrad optimizer with learning rate of $1\text{e-}4$. 


\subsection{Effect of Co-teaching}

To determine the effects of weak labels and the added value of co-teaching, we design an experiment comparing EDL with co-teaching to EDL alone. Table~\ref{coteaching_table} shows a promising improvement in both ECE score and patch-based balanced accuracy (Patch B-accuracy) when the co-teaching is employed. We report sensitivity, specificity and area under the curve (AUC) metrics for cores. Counter-intuitively, we observe that gains in patch-wise accuracy with co-teaching are not reflected in these metrics. We hypothesize that the averaging from patch-wise to core-wise predictions may sufficiently smooth the effects of noisy labels at this level. We emphasize that the AUC for \emph{both} methods is at least $10\%$ higher than AUC achieved using conventional ultrasound machines \cite{javadi2021training}, underlining the strong capabilities of high-frequency ultrasound.

\begin{figure}[tb]
\centering
\includegraphics[width=1.\linewidth]{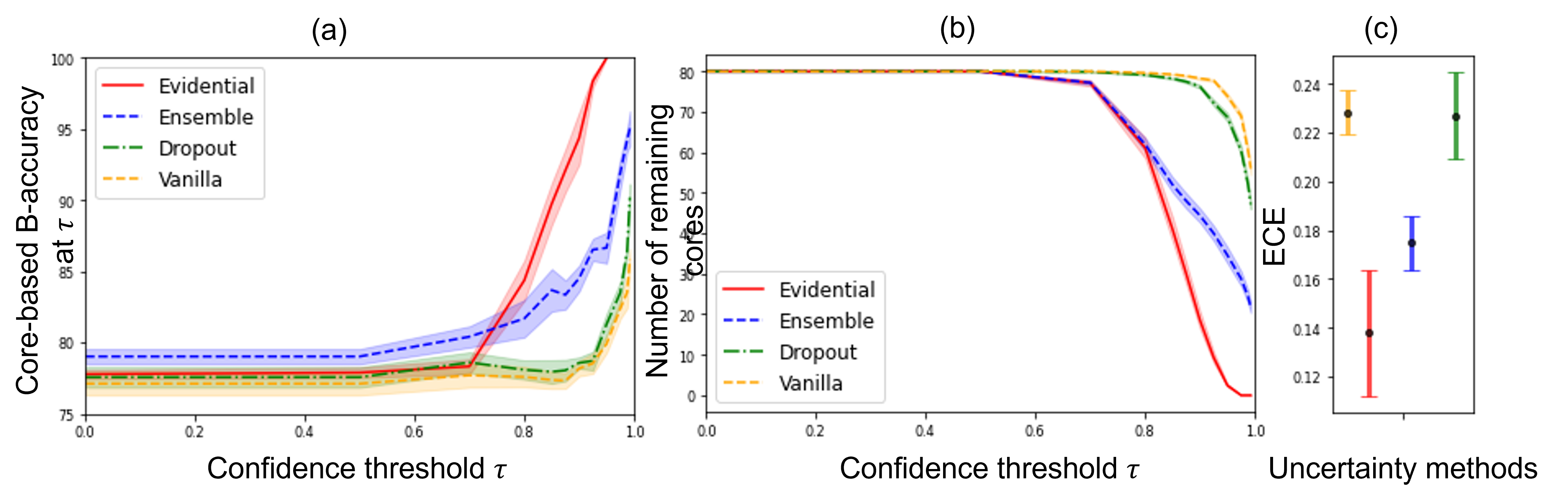}%

\caption{Left: accuracy vs. confidence plot. As we increase the confidence threshold $\tau$ and retain only confident predictions, the balanced accuracy increases accordingly. Middle: The number of remaining cores following exclusion based on the confidence threshold. Right: the Expected Calibration Error (ECE) error bar plot for all presented uncertainty quantification methods (lower is better).}
\label{fig_acc_number}
\end{figure}
\subsection{Comparison of Uncertainty Methods}
Quantification of predictive uncertainty could help clinical decision making during the biopsy procedure by only relying on highly confident predictions and discarding OOD and suspect samples. We examine EDL predictive uncertainty using \textit{accuracy vs. confidence plots} in this section, and illustrate how it may be utilised to eliminate uncertain predictions while achieving high accuracy on the confident ones. 
Then, we compare EDL predictive uncertainty with MC Dropout \cite{gal2016dropout} and deep ensemble \cite{lakshminarayanan2017simple} methods.

In our \textit{accuracy vs. confidence plot}, Figure~\ref{fig_acc_number} (a), we plot core-based balanced accuracy as a function of the confidence threshold $\tau \in [0,1]$ used to filter out underconfident patch-level predictions. Patches with predicted confidence less than $\tau$, i.e. predictive uncertainty more than $1-\tau$, are discarded. If at least 60\% of extracted patches for a biopsy core remain, the average of the remaining patch predictions is used as core-based prediction. 
We observe the increase in core-based accuracy as the threshold increases, showing that confident predictions tend to be correct. As shown in Figure~\ref{fig_acc_number} (b), there is a natural trade-off, with increased threshold values also resulting in increased numbers of rejected cores, yet with well-calibrated uncertainty methods it is not necessary to discard a high fraction of cores in order for uncertainty thresholding to result in meaningful accuracy gains. In Figure~\ref{fig_acc_number} (c), we compare the quality of predictive uncertainty of all methods via ECE score. Our experiments show that EDL achieves the best calibration error while providing the best balance between high accuracy and core retention at different threshold levels.

\subsection{Model Demonstration}
\begin{figure}[tb]
    \centering
    \includegraphics[scale=.40]{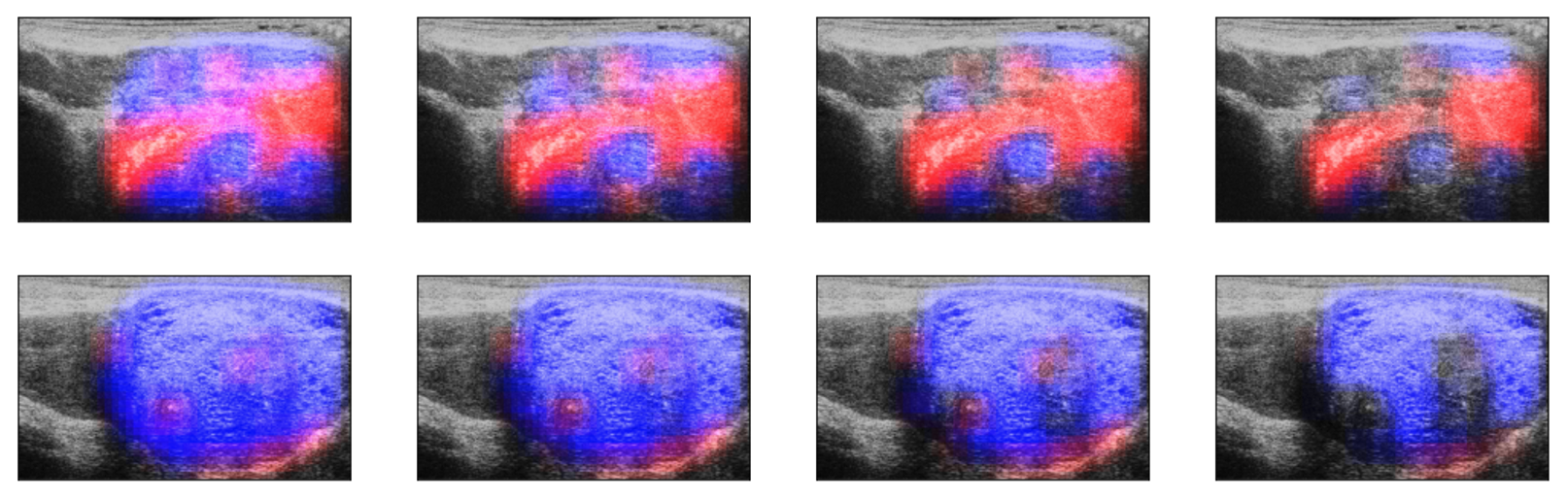}
    \caption{Heatmaps representing predictions of cancer (red) or benign (blue). The confidence threshold is increased from left to right as $[0.7, 0.8, 0.85, 0.9]$, progressively excluding more uncertain predictions. The top row is from cancerous core with Gleason score 4+3; the bottom row is from a benign core.}
    \label{fig:heatmaps}
\end{figure}
As a proof-of-concept for the clinical utility of our method, we applied our model as a sliding window over entire RF images and generated a heatmap, where red corresponds to a prediction of cancer and blue to a prediction of benign. Uncertainty thresholds at various levels were applied to discard uncertain predictions - discarded predictions had their opacity decreased to 0. These maps were overlaid over the corresponding B-mode images to visualize the spread of cancer. An example of heatmaps for a cancerous and benign core are shown in Figure~\ref{fig:heatmaps}. The cancerous image shows a large amount of red which focuses on two main regions as the confidence threshold increases. By the results of Figure~\ref{fig:my_label}, we can say that these loci are very likely to be cancerous lesions and good biopsy targets. The benign image, on the other hand, shows a dominance of blue, with two small red areas that disappear as the threshold increases. These are most likely areas of OOD features on which the model correctly reported high levels of uncertainty. These images show the subjective quality of our model's performance and the utility of an adjustable uncertainty threshold.

\section{Conclusion}
We proposed a model for confident PCa detection using micro-ultrasound. We employed co-teaching to improve robustness to label noise, and used evidential deep learning to model the predictive uncertainty of the model. We find these strategies to yield a significant improvement over baseline in the clinically relevant metrics of accuracy vs. confidence. Our model provides crucial confidence information to interventionists weighing the recommendations of the model against their own expertise, which can be critical for the adoption of precision biopsy targeting using TRUS. 

\subsubsection*{Acknowledgement.}
This work was supported by the Natural Sciences and Engineering Research Council of Canada (NSERC) and the Canadian Institutes of Health Research (CIHR).

\bibliographystyle{splncs04}
\bibliography{paper2769}
%




\end{document}